\documentclass[
amsmath,
amssymb,
aps,
prb,
reprint,
superscriptaddress,
longbibliography
]{revtex4-2}

\usepackage{mathtools}
\usepackage{multirow}
\usepackage{graphicx}
\usepackage{comment}
\usepackage{makecell}
\usepackage{braket}
\usepackage{diagbox}
\usepackage{bbding}
\usepackage[normalem]{ulem}
\usepackage{dcolumn}
\usepackage{bm}

\usepackage[dvipsnames]{xcolor}
\definecolor{darkblue}{HTML}{1B436C}

\usepackage[colorlinks=true]{hyperref}
\hypersetup{
    allcolors=darkblue
}

\newcommand{\affA}{State Key Laboratory of Artificial Microstructure and Mesoscopic
Physics, School of Physics, Peking University, Beijing 100871, China}

\begin{document}

\title{High-order resonances and emergent coherent dynamics in interacting Stark chains}

\author{Jiahe Chang}
 \affiliation{\affA}
 	\author{Hongzheng Zhao}
	\email{hzhao@pku.edu.cn}
 \affiliation{\affA}
	\date{\today}

\begin{abstract}

A linear potential gradient can strongly suppress quantum dynamics, while interactions can activate resonant many-body processes whose higher-order structure remains largely unexplored. Here we uncover a systematic hierarchy of high-order interaction--tilt resonances in an interacting Stark chain, where the resonance order counts the minimal number of elementary exchange steps involved. These resonances manifest as long-lived coherent oscillations with order-dependent timescales. By isolating individual resonant channels and developing a microscopic effective description, we connect these dynamical timescales to the underlying many-body exchange processes and reveal a pronounced dependence on the initial-state structure. Our results provide crucial insights into high-order many-body resonances and show how commensurate interaction-to-tilt ratios generate a hierarchy of parametrically slow coherent many-body dynamics.

\end{abstract}

    \maketitle
\let\oldaddcontentsline\addcontentsline
\renewcommand{\addcontentsline}[3]{}

\section{Introduction}

A linear gradient of potential energy can strongly suppress quantum dynamics in lattice systems. A paradigmatic example is Wannier--Stark localization, where a single particle localizes on a lattice under a uniform field~\cite{wannier1960wave,wannier1962dynamics}. In interacting many-body systems, this suppression can persist and give rise to long-lived nonergodic dynamics~\cite{schulz2019stark,van2019bloch,doggen2021stark,morong2021observation,guo2021stark,wang2021stark}. In the strong-tilt regime, the residual dynamics can be governed by emergent dipole conservation and correlated pair-hopping, closely connected to Hilbert-space fragmentation~\cite{sala2020ergodicity,khemani2020localization,moudgalya2022thermalization,burin2022exact,kim2025multi}.

A qualitatively different situation arises when the interaction strength becomes comparable to the potential tilt. Interaction--tilt resonances can then occur when the potential-energy change associated with a dynamical process, such as particle hopping or spin exchange, is compensated by the corresponding interaction-energy change, thereby activating processes that would otherwise be strongly off-resonant.

Different resonant channels can, however, require different numbers of elementary tunneling events to connect the relevant configurations. We therefore define the resonance order $n$ as this minimal number of tunneling steps. Low-order resonances are crucial for a variety of nontrivial phenomena in tilted Bose- and Fermi-Hubbard systems, including quantum phase transitions within resonant subspaces~\cite{sachdev2002mott,simon2011quantum,buyskikh2019spin}, Hilbert-space fragmentation~\cite{scherg2021observing,kohlert2023exploring}, and many-body scarring~\cite{desaules2021proposal,su2023observation}. Higher-order resonances have also been observed in cold-atom systems~\cite{meinert2014observation,zhang2026scalable}. What remains less understood is how to systematically classify high-order resonances and develop a microscopic description of the processes underlying their coherent many-body dynamics. 

Systematically characterizing these high-order processes is challenging for several reasons. When the tunneling amplitude is smaller than the relevant diagonal energy scales, the effective higher-order coupling is strongly suppressed. Its characteristic timescale therefore grows rapidly with resonance order, making high-order dynamics susceptible to being masked by lower-order processes or unresolved within accessible observation times. Further, generic many-body configurations can also support multiple competing resonant processes, making individual high-order channels difficult to isolate. Finally, operator-level effective descriptions, well established for low-order resonances~\cite{sachdev2002mott,buyskikh2019spin,scherg2021observing,kohlert2023exploring}, become increasingly involved at higher orders, complicating both the identification of microscopic resonant processes and a quantitative description of their dynamics.

In this work, we overcome these difficulties by studying the quench dynamics of an interacting Stark spin chain starting from a single-domain-wall initial state. The domain wall provides a spatially isolated active region that strongly reduces the number of competing resonant processes, such that individual high-order resonant channels can be resolved. In this setting, we uncover a systematic hierarchy of high-order resonances accompanied by long-lived coherent oscillations. By combining energy-matching conditions with the microscopic constraints imposed by nearest-neighbor spin exchange, we determine the minimal order of each resonant channel and thereby predict the scaling of its characteristic timescale. These predictions are verified numerically, and for the third-order resonance we derive an explicit effective Hamiltonian that identifies the underlying microscopic process and quantitatively reproduces the full-model dynamics. 

Our results reveal how commensurate interaction-to-tilt ratios generate a hierarchy of parametrically slow coherent many-body resonances.

\section{Setup}

We consider an even-length spin-$1/2$ chain of $L$ sites subject to a linear (Stark) potential, with nearest-neighbor interactions and open boundary conditions. This chain is described by the Hamiltonian
\begin{equation}
    H{=}\!\sum_{i=1-L/2}^{L/2-1}\!\left[J(X_iX_{i+1}+Y_iY_{i+1}){+}\Delta Z_iZ_{i+1}\right]{+}W\!\sum_{i}\!iZ_i ,
    \label{eq:Hspin}
\end{equation}
where $X_{i}$, $Y_{i}$, and $Z_{i}$ denote Pauli matrices acting on site $i$. We apply a linear potential with gradient $W$, giving a site-dependent strength $\epsilon_{i} = Wi$. $\Delta$ sets the strength of the nearest-neighbor Ising interaction. Throughout this work we set $J=\hbar=1$ and both $W$ and $\Delta$ are taken to be positive. This Hamiltonian conserves the total magnetization along the $z$ direction.

We choose a pure product state with zero total magnetization featuring a single domain wall. All spins on the left half of the chain ($i\le 0$) are polarized up, whereas all spins on the right half ($i>0$) are polarized down. Explicitly,
\begin{equation}
    \left | \psi_0  \right \rangle = |\underbrace{\uparrow \uparrow \cdots \uparrow \uparrow \uparrow }_{L/2} \, \underbrace{\downarrow \downarrow \downarrow \cdots \downarrow \downarrow }_{L/2} \rangle \, .
    \label{eq:initial_state}
\end{equation}
For $W > 0$, this configuration minimizes the linear-potential energy among all zero-magnetization product states.

\section{Dynamics}

We quench the single-domain-wall initial state $|\psi_{0}\rangle$ with the static Hamiltonian~\eqref{eq:Hspin} at moderate $W$ and over a wide range of $\Delta$. For $\Delta=0$, the model maps to free fermions and can be understood in terms of Wannier--Stark localization, for which the single-particle localization length scales inversely with the field gradient, i.e., $\xi_{\mathrm{sp}} \propto 1/W$~\cite{wannier1960wave,wannier1962dynamics}. Upon introducing weak interactions, localization can persist even in the absence of disorder~\cite{schulz2019stark,van2019bloch,doggen2021stark}. In the strongly interacting limit $\Delta \gg W,J$, the Hamiltonian~\eqref{eq:Hspin} is dominated by the Ising interaction. Consequently, $z$-polarized product states are approximate eigenstates, and the quench dynamics are strongly suppressed.

At intermediate interaction strengths, however, the initially sharp domain wall can melt during time evolution. The extent of this melting provides a direct measure of localization~\cite{hauschild2016domain}: substantial domain-wall melting is characterized by a wide region in which the local magnetization deviates appreciably from its initial value, whereas localized dynamics restrict such deviations to only a few sites around the initial domain wall. Interestingly, near certain simple fractional values of the ratio $\Delta/W$, we observe long-period coherent dynamics rather than simple domain-wall melting.

\begin{figure}[t]
    \centering
    \includegraphics[width=\linewidth]{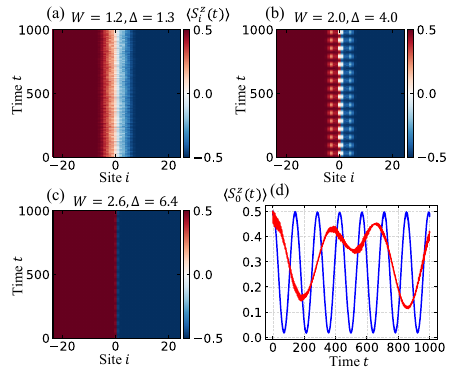}
    \caption{Quench dynamics for different parameters. 
    (a)--(c) Spatiotemporal maps of $\langle S_{i}^{z}(t) \rangle$ for three representative parameter points, characterized respectively by a smooth transition region, coherent oscillations, and a sharp domain wall. (d) Time traces of $\langle S_{0}^z(t)\rangle$ for two cases exhibiting coherent oscillations. The blue curve corresponds to $W=8.0$, $\Delta=12.0$, while the red curve corresponds to $W=2.0$, $\Delta=5.0$. Although both exhibit long-lived coherent oscillations, their temporal structures are different: the former is nearly cosine-like, whereas the latter shows a more complex oscillatory pattern.}
    \label{fig:quench_dynamics}
\end{figure}

To show these distinct dynamical behaviors, we first present the spatiotemporal profiles of $\langle S_i^z(t)\rangle=\langle Z_i(t)\rangle/2$ in Fig.~\ref{fig:quench_dynamics}(a)--(c), with red and blue denoting positive and negative magnetization, respectively. {The dynamical results are obtained using the time-evolving block decimation (TEBD) algorithm, an MPS-based method well suited to the present Stark-localized dynamics, where entanglement spreading remains sufficiently slow over the simulated time scales, allowing access to system sizes well beyond exact diagonalization (ED)~\cite{hauschild2018efficient}.} We use $L=48$ and evolve the system up to $T=1000$. In panel~(a), where $W=1.2$ and $\Delta=1.3$, the initially sharp domain wall gradually melts, resulting in a smooth spatial profile with only weak fluctuations at long times. A broad transition region, visible as the light-colored region around the chain center, connects the two fully polarized domains. By contrast, panel~(c), corresponding to $W=2.6$ and $\Delta=6.4$, exhibits much more strongly suppressed dynamics. The initial domain wall remains sharp throughout the simulation, and appreciable changes in magnetization are confined to only a few spins immediately adjacent to it.

\begin{figure*}[t]
    \centering
    \includegraphics[width=0.9\textwidth]{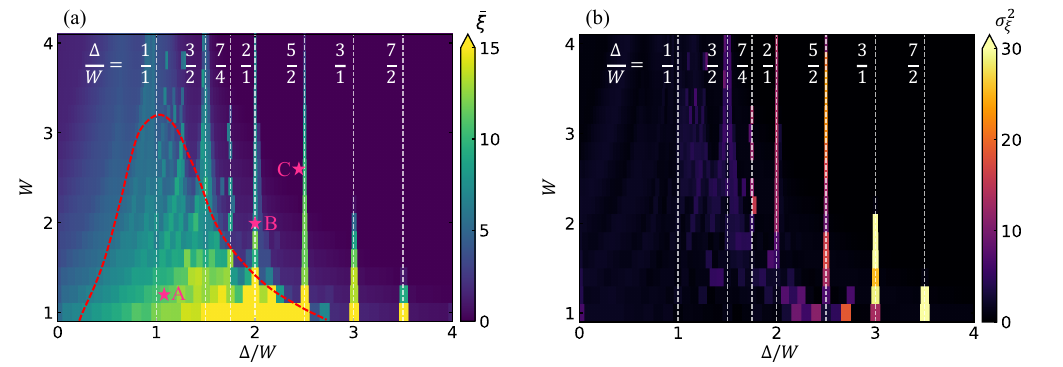}
    \caption{Localization-length maps. Colormaps of (a) the time-averaged localization length $\bar{\xi}$ and (b) its temporal fluctuations $\sigma_{\xi}^{2}$ as functions of $W$ and $\Delta/W$, obtained from TEBD simulations for $L=48$ up to $T=1000$. (a) Brighter colors indicate a larger average extent of domain-wall melting. The red dashed curve marks the contour $\bar{\xi}=10$ as a guide, separating the broad-melting region at smaller $W$ from the strongly suppressed melting at larger $W$. White dashed lines mark simple fractional values of $\Delta/W$ where $\bar{\xi}$ is anomalously enhanced. Points $A$, $B$, and $C$ correspond to Fig.~\ref{fig:quench_dynamics}(a)--(c), respectively. (b) Brighter colors indicate stronger temporal fluctuations. Enhanced $\sigma_{\xi}^{2}$ is concentrated near the same fractional values of $\Delta/W$ marked by the white dashed lines in panel~(a), consistent with the coherent oscillations shown in Fig.~\ref{fig:quench_dynamics}(b) and (d). Together, the two maps reveal narrow resonant structures superimposed on the broader dependence of domain-wall melting on $W$ and $\Delta/W$.}
    \label{fig:LLL_map}
\end{figure*}

Panel~(b), however, displays a remarkably different and highly structured spatiotemporal pattern. For $W=2.0$ and $\Delta=4.0$, the domain-wall region undergoes slow, coherent oscillations with a period much longer than the typical time scale of local fluctuations in panel~(a). Similar long-lived coherent oscillations are also observed for other parameter choices, although their temporal structures can differ substantially. Figure~\ref{fig:quench_dynamics}(d) illustrates this difference through the $z$-magnetization at the site immediately to the left of the initial domain wall ($i=0$), where the oscillatory dynamics are most pronounced. For $W=8.0,\Delta=12.0$ shown as the blue curve, $\langle S^{z}_{0}(t)\rangle$ exhibits a nearly cosine-like oscillation, whereas for $W=2.0,\Delta=5.0$ shown as the red curve, the oscillatory pattern is more complex.

The above observations reveal three qualitatively distinct dynamical behaviors, characterized respectively by a broad and smooth transition region, coherent oscillations, and strongly suppressed dynamics. To quantify these behaviors, we introduce a \textit{localization length} that measures the spatial extent of domain-wall melting. Fixing a threshold $\varepsilon=0.05$, we define the left and right boundaries of the melted region at time $t$ as
\[
\begin{split}
\ell(t) &= \min\{\, i \,:\, \langle S_i^z(t)\rangle < +\tfrac{1}{2}-\varepsilon \,\}, \\ 
r(t) &= \max\{\, i \,:\, \langle S_i^z(t)\rangle > -\tfrac{1}{2}+\varepsilon \,\}.
\end{split}
\]
The localization length is then defined as
\begin{equation}
\xi(t)=r(t)-\ell(t)+1.
\label{localization_length}
\end{equation}
We further characterize its average extent and temporal fluctuations by
\begin{equation}
\bar{\xi}=\frac{1}{T}\int_0^T \xi(t)\,dt,\qquad
\sigma_{\xi}^{2}=\frac{1}{T}\int_0^T[\xi(t)-\bar{\xi}]^2\,dt.
\label{eq:xi_statistics}
\end{equation}
These two quantities provide complementary information: strongly confined dynamics are characterized by a small $\bar{\xi}$, whereas pronounced oscillations of the domain-wall region give rise to an enhanced $\sigma_{\xi}^{2}$.

Fig.~\ref{fig:LLL_map}(a) displays the time-averaged localization length $\bar{\xi}$, with colors encoding its magnitude. Bright colors, corresponding to large $\bar{\xi}$, are mainly found in the lower part of the plot, i.e., for smaller $W$ and intermediate values of $\Delta/W$. The parameter region below the red dashed curve, which approximately marks $\bar{\xi}=10$, corresponds to the dynamics shown in Fig.~\ref{fig:quench_dynamics}(a), where simple melting of the initial domain wall produces a smooth magnetization profile and hence a large $\bar{\xi}$. By contrast, the dark-colored region, typically located above the red dashed curve, corresponds to strongly suppressed domain-wall melting. There, $\xi(t)$ remains small throughout the evolution, consistent with the nearly frozen dynamics shown in Fig.~\ref{fig:quench_dynamics}(c). Both types of dynamics exhibit weak temporal fluctuations of the localization length, as reflected by the relatively small values of $\sigma_{\xi}^{2}$ in Fig.~\ref{fig:LLL_map}(b).

A qualitatively different behavior appears along several narrow structures above the red dashed curve in Fig.~\ref{fig:LLL_map}(a), where $\bar{\xi}$ is anomalously enhanced at specific fractional values of $\Delta/W$, as indicated by the white dashed lines. These structures become even more pronounced in the temporal-fluctuation map in Fig.~\ref{fig:LLL_map}(b), where $\sigma_\xi^2$ is strongly enhanced at the same parameter values. Together with the coherent oscillations observed in Fig.~\ref{fig:quench_dynamics}(b) and (d), these features provide a clear signature of resonant dynamics. We note that the apparent termination of these structures at large $W$ is a finite-time effect: as the oscillations become progressively slower, their signatures can no longer be fully resolved within the simulation time $T$, leading to reduced values of both $\bar{\xi}$ and $\sigma_\xi^2$.

\section{Resonance}

The coherent oscillations originate from resonant processes in which the changes in Stark and interaction energies compensate each other. Crucially, we determine the minimal number of spin-exchange steps required to connect resonant configurations, which in turn fixes its perturbative order and characteristic timescale.

For a Fock state in the $z$ basis, the Stark-potential contribution to the diagonal energy is
\begin{equation}
E_{\mathrm{pot}}=W\sum_{i=1-L/2}^{L/2}i\langle Z_i\rangle=2W\sum_{i\in\uparrow}i+\mathrm{Const},
\end{equation}
where $\sum_{i\in\uparrow}$ denotes a sum over all up-spin sites, $\langle Z_i\rangle=\pm1$, and the constant is state independent within the zero-magnetization sector. The Ising contribution is
\begin{equation}
E_{\mathrm{ZZ}}=\Delta\sum_{i=1-L/2}^{L/2-1}\langle Z_iZ_{i+1}\rangle=\Delta(L-1-2N_d),
\end{equation}
where $N_d$ denotes the number of domain walls in the Fock state.

To quantify how this diagonal energy changes between an initial and a final Fock state connected by a sequence of spin exchanges, we define
\[
P=\sum_{i\in\uparrow,\mathrm{final}}i-\sum_{i\in\uparrow,\mathrm{initial}}i,\qquad Q=N_{d,\mathrm{final}}-N_{d,\mathrm{initial}},
\]
where $P$ measures the net change in the sum of up-spin positions and $Q$ measures the change in the number of domain walls. The corresponding changes in the diagonal energy are
\begin{equation}
\delta E_{\mathrm{pot}}=2WP,\qquad \delta E_{\mathrm{ZZ}}=-2\Delta Q.
\end{equation}
A resonant process requires the diagonal-energy mismatch between the connected Fock states to vanish due to energy conservation, which means $\delta E_{\mathrm{pot}}+\delta E_{\mathrm{ZZ}}=0$, yielding
\[
\frac{\Delta}{W}=\frac{P}{Q}.
\]
Both $P$ and $Q$ are integers by definition, exact diagonal-energy matching therefore occurs only at fractional values of $\Delta/W$. We parameterize such a value as
\begin{equation}
\frac{\Delta}{W}=\frac{p}{q},
\label{eq:resonance_condition}
\end{equation}
where $p$ and $q$ are coprime positive integers.

\begin{figure}[t]
    \centering
    \includegraphics[width=\linewidth]{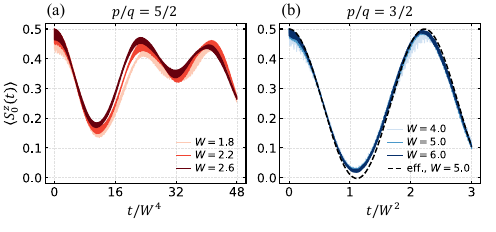}
    \caption{Time-rescaled collapse of resonant dynamics. At fixed $\Delta/W=p/q$, the local magnetization $\langle S_{0}^{z}(t)\rangle$ for different $W$ approximately collapses under the rescaling $\tau=t/W^{n_{\min}-1}$. Panels~(a) and (b) correspond to $p/q=5/2$ and $3/2$, with $n_{\min}=5$ and $3$, respectively. The black dashed curve in panel~(b) is obtained from the third-order effective Hamiltonian described below and closely reproduces the full-model dynamics. The collapse confirms that $n_{\min}$ sets the characteristic dynamical timescale; residual deviations arise from subleading higher-order corrections.}
    \label{fig:overlap_dynamics}
\end{figure}

For the initial state~\eqref{eq:initial_state}, $N_{d,\mathrm{initial}}=1$. Since this state minimizes $\sum_{i\in\uparrow}i$ within the zero-magnetization sector, any nontrivial resonant final state must have $P>0$ and $Q>0$. Away from the open boundaries, each nearest-neighbor spin exchange changes the number of domain walls by $0$ or $\pm2$, so any Fock state reached from the initial state has $Q=2m$ with $m\in\mathbb{N}^{+}$. Boundary processes do not affect the dynamics considered here.

Writing the resonance ratio as Eq.~\eqref{eq:resonance_condition} and $Q$ as $2m$, the condition $P/Q=p/q$ gives $P=2mp/q$. Since $P$ is an integer and $p$ and $q$ are coprime, $q$ must divide $2m$. The smallest allowed $m$ is therefore $q$ for odd $q$ and $q/2$ for even $q$, yielding $P_{\min}=2p$ and $P_{\min}=p$, respectively. Because each elementary spin exchange changes $P$ by only one, any resonant process requires at least $P_{\min}$ exchange steps. For the single-domain-wall configuration, this lower bound is attainable, giving the minimal resonance order as
\begin{equation}
n_{\min}=
\begin{cases}
2p, & q\ \mathrm{odd},\\
p, & q\ \mathrm{even}.
\end{cases}
\label{eq:resonance_order}
\end{equation}
Thus, the fractional ratio $\Delta/W=p/q$ determines both the diagonal-energy matching condition and the minimal perturbative order of the corresponding resonant process.

We note that low-order interaction--tilt resonances have been explored previously in related tilted Hubbard systems~\cite{scherg2021observing,kohlert2023exploring,desaules2021proposal,su2023observation}. These low-order cases, including the first-order resonance at $\Delta/W=1/2$ and the second-order resonance at $\Delta/W=1$ in our notation, are discussed further in Appendix~\ref{app:low_order}. In the main text, we go beyond these low-order resonances and uncover a hierarchy of higher-order resonant processes with parametrically small effective couplings, leading to long-lived and coherent many-body oscillations.

Concretely, from perturbation theory, as discussed further in the following section, an $n$th-order process generated by the exchange term has an effective coupling strength $J_{\mathrm{eff}}\propto J^{n}/W^{n-1}$. Accordingly, at a fixed resonance ratio $\Delta/W$, the characteristic frequency of the coherent oscillations in the local magnetization is expected to scale as $\omega\propto J^{n}/W^{n-1}$. We numerically verify this scaling in Fig.~\ref{fig:overlap_dynamics}, where the time traces of the local magnetization $\langle S_{0}^{z}(t)\rangle$ for different values of $W$ collapse onto each other when plotted as a function of the rescaled time $\tau\equiv t/W^{n_{\min}-1}$. Small deviations from a perfect collapse arise because the parameters considered are only moderately larger than $J$ and therefore do not lie deep in the perturbative regime $W,\Delta\gg J$, leaving subleading higher-order corrections non-negligible. These deviations are systematically reduced as $W$ is increased.

\begin{figure}[t]
    \centering
    \includegraphics[width=\linewidth]{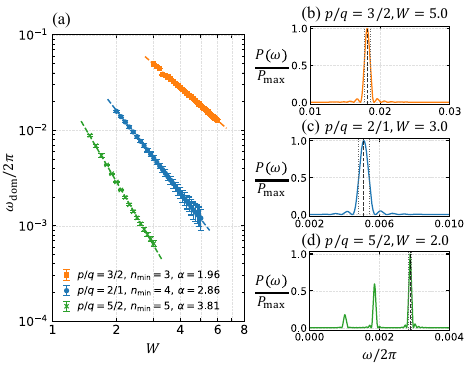}
    \caption{Dominant-frequency scaling of the resonant dynamics. (a) Dominant oscillation frequency $\omega_{\mathrm{dom}}/2\pi$, extracted from the frequency spectrum of $\langle S_{0}^{z}(t)\rangle$, as a function of the Stark gradient $W$ on log--log axes for resonance ratios $p/q=3/2$, $2$, and $5/2$. The error bars indicate the lower and upper half-maximum frequencies of the dominant spectral peak. Power-law fits $\omega_{\mathrm{dom}}\sim W^{-\alpha}$ yield exponents consistent with the prediction $\alpha=n_{\min}-1$. (b)--(d) Representative frequency spectra for $(W,\Delta)=(5.0,7.5)$, $(3.0,6.0)$, and $(2.0,5.0)$, corresponding to $p/q=3/2$, $2$, and $5/2$, respectively. The corresponding time windows are $[0,1000]$, $[0,1500]$, and $[0,10000]$. Each spectral power $P(\omega)$ is normalized by its maximum value $P_{\max}$.}
    \label{fig:dominant_frequency}
\end{figure}

As a complementary test of the predicted resonance order, we extract the dominant oscillation frequency $\omega_{\mathrm{dom}}$ from the frequency spectrum of the local magnetization $\langle S_{0}^{z}(t)\rangle$, with representative spectra shown in Fig.~\ref{fig:dominant_frequency}(b)--(d). Figure~\ref{fig:dominant_frequency}(a) shows $\omega_{\mathrm{dom}}$ as a function of $W$ for three representative resonance ratios $p/q=3/2$, $2$, and $5/2$, whose predicted minimal resonance orders are $n_{\min}=3,4,$ and $5$, respectively. In all three cases, the dominant frequency follows a power-law scaling
\[
\omega_{\mathrm{dom}}\sim W^{-\alpha},
\]
with the fitted exponent $\alpha$ in agreement with the prediction $\alpha=n_{\min}-1$ from Eq.~\eqref{eq:resonance_order}.

These coherent oscillations are remarkably long-lived. No discernible damping is observed even when the simulation time is extended to $T\sim10^{5}$. This behavior is associated with the restricted resonant subspace explored by the initial state $|\psi_0\rangle$, which is formed by Fock states satisfying the diagonal-energy matching condition. Importantly, the dimension of this subspace saturates with increasing system size $L$, as shown in Appendix~\ref{app:subspace_size}. The dynamics therefore remain effectively confined to a finite-dimensional resonant manifold, strongly suppressing thermalization and becoming essentially insensitive to $L$ once the system is sufficiently large.

To resolve the microscopic processes governing the dynamics within this resonant subspace, we now derive an effective Hamiltonian.

\section{Effective Hamiltonian}

We now construct an effective description of the resonant dynamics. At commensurate interaction-to-tilt ratios, a transformation to the rotating frame generated by the diagonal Stark and interaction terms yields a time-periodic Hamiltonian, which can be treated systematically using the van Vleck high-frequency expansion~\cite{bukov2015universal,eckardt2015high,sen2021analytic}. This construction applies generally to commensurate resonances; below we focus on the third-order resonance at $\Delta/W=3/2$.

To derive the effective Hamiltonian explicitly, we first rewrite Hamiltonian~\eqref{eq:Hspin} in terms of spinless fermions through a Jordan--Wigner transformation~\cite{jordan1928paulische}. Identifying an up spin with an occupied site and a down spin with an empty one gives $Z_i=2n_i-1$. The spin-exchange term maps onto nearest-neighbor fermion hopping, while the Stark potential and Ising interaction become a linear potential and a nearest-neighbor density interaction, respectively. Up to a state-independent constant and boundary terms that do not affect the bulk resonant processes, Hamiltonian~\eqref{eq:Hspin} becomes
\begin{equation}
H_{\mathrm{fermion}}=\mathsf{t}\sum_i(c_i^{\dagger}c_{i+1}+\mathrm{h.c.})+F\sum_i i n_i+U\sum_i n_i n_{i+1},
\label{eq:Hfermion}
\end{equation}
with $\mathsf{t}=2J$, $F=2W$, and $U=4\Delta$. In particular, the third-order resonance $\Delta/W=3/2$ corresponds to $U/F=3$ in the fermionic representation. We transform to the rotating frame generated by the diagonal Stark and interaction terms,
\begin{equation}
V(t)=e^{-it\left(F\sum_i i n_i+U\sum_i n_i n_{i+1}\right)}, 
\end{equation}
such that the Hamiltonian in this frame becomes
\begin{equation}
\begin{split}
H^{\mathrm{rot}}(t)&=V^{\dagger}(t)H_{\mathrm{fermion}}V(t)-iV^{\dagger}(t)\partial_tV(t)\\
&=\mathsf{t}\sum_i e^{-iFt}e^{iUt(n_{i-1}-n_{i+2})}c_i^{\dagger}c_{i+1}+\mathrm{h.c.}.
\end{split}
\label{eq:Hrot}
\end{equation}
Thus, the diagonal energy differences that determined the resonance condition now appear as time-dependent phases of the fermion-hopping processes.

Since $n_i$ is a projector satisfying $n_i^2=n_i$, the density-dependent phase factor in Eq.~\eqref{eq:Hrot} can be decomposed according to the occupations of the two neighboring sites as
\[
\begin{split}
    e^{iUt(n_{i-1}-n_{i+2})}
    =
    \, (1-n_{i-1})(1-n_{i+2})
    +
    n_{i-1}n_{i+2} \\
    +
    e^{iUt}n_{i-1}(1-n_{i+2})+
    e^{-iUt}(1-n_{i-1})n_{i+2}.
\end{split}
\]
Consequently, depending on the occupations of sites $i-1$ and $i+2$, the hopping term in $H^{\mathrm{rot}}(t)$ acquires a phase that oscillates at different frequencies $\pm F$, $\pm(F-U)$, and $\pm(F+U)$. For the third-order resonance considered here ($U/F=3$), these frequencies become $\pm F$, $\pm2F$, and $\pm4F$. The rotating-frame Hamiltonian is therefore periodic with fundamental frequency $F$ and can be written as
\begin{equation}
    H^{\mathrm{rot}}(t) = \sum_{l\in\mathbb{Z}} H_l e^{ilFt},
\label{eq:Fourier_component}
\end{equation}
with nonzero Fourier components $H_{\pm1}$, $H_{\pm2}$, and $H_{\pm4}$.

In the regime $F \gg \mathsf{t}$ and with the resonance ratio $U/F=3$ kept fixed, the rapid oscillations at frequencies $F$ and its integer multiples can be perturbatively eliminated. The remaining slow resonant dynamics can then be described by a time-independent effective Hamiltonian. Since the Fourier components $H_l$ are of order $\mathsf{t}$, this effective Hamiltonian can be systematically expanded in powers of the small parameter $\mathsf{t}/F$ using the van Vleck high-frequency expansion~\cite{bukov2015universal}. Up to $1/F^2$, the expansion reads
\begin{equation}
\begin{split}
    &H_{\mathrm{eff}}^{(0)} = H_{0}, \qquad H_{\mathrm{eff}}^{(1)} = \frac{1}{F} \sum_{l=1}^{\infty}\frac{1}{l} \left[H_l, H_{-l} \right] ,\\
    &H_{\mathrm{eff}}^{(2)} {=} \frac{1}{F^2} {\sum_{l \ne 0}}\Bigg(\frac{[H_{-l}{,}[H_0{,}H_l]]}{2l^2} {+} {\sum_{l^{\prime} {\ne} 0, l}} \frac{[H_{-l^{\prime}}{,} [H_{l^{\prime}-l}{,} H_l]]}{3ll^{\prime}}\Bigg){.}
    \label{eq:HFE}
\end{split}
\end{equation}
Because Fourier component $H_0=0$ at $U/F=3$, the contribution $H_{\mathrm{eff}}^{(0)}$ vanishes, while both $H_{\mathrm{eff}}^{(1)}$ and $H_{\mathrm{eff}}^{(2)}$ remain relevant. The former contains diagonal corrections together with second-order off-diagonal dynamical terms with characteristic scale $\mathsf{t}^2/F$, whereas the latter contains third-order dynamical terms with characteristic scale $\mathsf{t}^3/F^2$. The complete operator expressions for the effective Hamiltonian and derivation details are given in Appendix~\ref{app:effective_Hamiltonian}.

Most of the off-diagonal dynamical terms generated in $H_{\mathrm{eff}}^{(1)}$ and $H_{\mathrm{eff}}^{(2)}$ give zero when acting on the initial single-domain-wall state~\eqref{eq:initial_state}. Only two density-dependent hopping processes,
\begin{equation}
\begin{split}
T_1& = \sum_i n_{i-1}(1-n_{i+1})(1-n_{i+2})(1-n_{i+4})\,c_{i+3}^{\dagger}c_i,\\
T_2& = \sum_i n_{i-1}n_{i+1}n_{i+2}(1-n_{i+4})\,c_{i+3}^{\dagger}c_i,
\end{split}
\label{eq:third_order_dynamical}
\end{equation}
connect the initial state to other Fock states as
\[
T_1|11110000\rangle{\propto}|11100010\rangle,\quad T_2|11110000\rangle{\propto}|10111000\rangle,
\]
where $1$ and $0$ denote occupied and empty fermionic sites, respectively. Although subject to different local occupation constraints, both processes transfer a fermion three lattice sites to the right. Consequently, the linear-potential energy increases by $3F$, while the nearest-neighbor interaction energy decreases by $U$, so the total diagonal-energy change vanishes at $U/F=3$. In the spin language, these terms correspond to an effective transfer of up-spin polarization across three lattice sites via three successive nearest-neighbor exchanges, providing a concrete microscopic explanation of the third-order resonance predicted by the energy-matching condition.

To test the resulting effective description directly, we quench the same initial state under the effective Hamiltonian at $F=10.0$ and $U=30.0$, corresponding to $W=5.0$ and $\Delta=7.5$ in the spin representation. Numerically, the effective Hamiltonian can be constructed either from the closed operator form derived above or, equivalently, from matrix elements obtained by summing virtual hopping paths within the resonant subspace; the latter formulation also generalizes naturally to higher-order resonances, as discussed in Appendix~\ref{app:effective_Hamiltonian}. The resulting local occupation dynamics are converted back to spin magnetization through $\langle S_0^z(t)\rangle=\langle n_0(t)\rangle-\tfrac{1}{2}$ and are shown by the black dashed curve in Fig.~\ref{fig:overlap_dynamics}(b). The effective-Hamiltonian dynamics closely reproduce those of the full model, including both the coherent oscillation pattern and its characteristic timescale, demonstrating that the effective description quantitatively captures the microscopic third-order resonant dynamics.

\section{Discussion}

We have uncovered a hierarchy of long-lived, high-order coherent many-body resonances. Their longevity appears closely related to the size of the energy-matched subspace accessible from the initial state. As shown in Appendix~\ref{app:subspace_size}, the number of Fock configurations with the same diagonal energy saturates with system size for the single-domain-wall state, but grows exponentially for a charge-density-wave state. This pronounced initial-state dependence suggests that the resonant dynamics may exhibit an emergent weak-fragmentation structure~\cite{sala2020ergodicity,moudgalya2022quantum,wang2025exploring}.

Our findings can be extended to states containing multiple domain walls. Well-separated walls should evolve approximately independently, suggesting a broader family of initial states that can support long-lived nonergodic dynamics. As the walls approach one another, or as the spatial extent of their resonant dynamics grows, different high-order channels can interact and proliferate. Whether this proliferation ultimately restores ergodicity remains an open question, conceptually reminiscent of the role of many-body resonances in destabilizing disorder-induced MBL~\cite{morningstar2022avalanches,ha2023many}, despite their distinct microscopic origins.

The resonance mechanism is not restricted to nearest-neighbor interactions or to one dimension. More general diagonal interactions can support analogous resonances whenever discrete interaction-energy changes compensate the Stark-energy cost, extending the mechanism to longer-range and higher-body interactions. In higher dimensions, the larger number of microscopic paths connecting energy-matched configurations may give rise to a richer variety of competing resonant channels~\cite{adler2024observation}.

The main ingredients of our setup---interacting dynamics in a strong potential gradient together with controlled preparation of spatially structured product states---are already available in state-of-the-art quantum simulators~\cite{morong2021observation,guo2021stark,adler2024observation}, placing the predicted resonance hierarchy within experimental reach.

\begin{acknowledgments}
We thank Andrea Pizzi, Biao Huang, and Marin Bukov for stimulating discussions. This work is supported by Quantum Science and Technology-National Science and Technology Major Project
(No. 2024ZD0301800) and by the National Natural Science
Foundation of China (Grant No. 12474214), and by ``The Fundamental Research Funds for the Central Universities, Peking University”, and by ``High-performance Computing Platform of Peking University".
\end{acknowledgments}

\appendix
\renewcommand{\appendixname}{APPENDIX}
\renewcommand{\thefigure}{\thesection\arabic{figure}}
\makeatletter
\@addtoreset{figure}{section}
\makeatother

\section{LOW-ORDER RESONANCES}
\label{app:low_order}

Low-order resonances, including the first-order resonance at $p/q=1/2$ and the second-order resonance at $p/q=1$ in our notation, have been extensively studied in tilted Hubbard models~\cite{sachdev2002mott,simon2011quantum,buyskikh2019spin,scherg2021observing,kohlert2023exploring,desaules2021proposal,su2023observation}, with corresponding effective descriptions developed in Refs.~\cite{sachdev2002mott,buyskikh2019spin,scherg2021observing,kohlert2023exploring}. To compare these low-order cases directly with the higher-order resonances discussed in the main text, we quench the same single-domain-wall initial state and extract the dominant oscillation frequencies from the frequency spectra of the local magnetization $\langle S_{0}^z(t)\rangle$ as a function of the Stark gradient $W$.

\begin{figure}[t]
    \centering
    \includegraphics[width=0.9\linewidth]{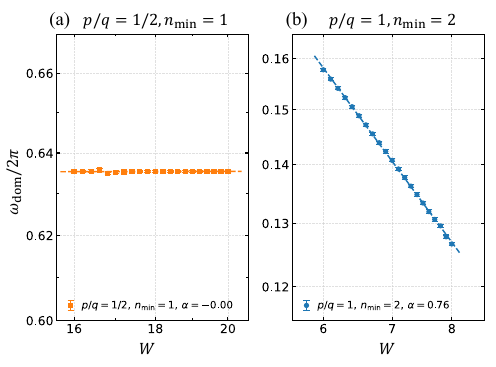}
    \caption{
    Dominant-frequency scaling of low-order resonances.
    The dominant oscillation frequency $\omega_{\mathrm{dom}}/2\pi$, extracted from the frequency spectrum of $\langle S_{0}^{z}(t)\rangle$, is shown as a function of the Stark gradient $W$ on log--log axes for (a) $p/q=1/2$ with $n_{\min}=1$, and (b) $p/q=1$ with $n_{\min}=2$.
    The spectra are obtained from the time window $t\in[0,1500]$.
    Dashed lines show power-law fits $\omega_{\mathrm{dom}}\sim W^{-\alpha}$.
    The fitted behavior is consistent with the expected order-dependent scaling $\alpha=n_{\min}-1$.
    }
    \label{fig:low_order_resonance}
\end{figure}

Figure~\ref{fig:low_order_resonance} shows the resulting dominant frequencies as a function of $W$ on log--log axes. For the first-order resonance, the dominant frequency is nearly independent of $W$, consistent with $\alpha=n_{\min}-1=0$. For the second-order resonance, the dominant frequency decreases with increasing $W$ and exhibits the predicted scaling $\alpha=n_{\min}-1=1$. Together with the higher-order resonances discussed in the main text, these results support the general scaling
\[
\omega_{\mathrm{dom}}\sim W^{-\alpha},
\qquad
\alpha=n_{\min}-1.
\]

Compared with the higher-order resonances discussed in the main text, resolving these low-order resonant oscillations requires substantially larger values of the Stark gradient $W$. At smaller $W$, the resonant dynamics are superimposed on a relatively strong background of domain-wall melting, making the coherent oscillations less clearly distinguishable. We therefore use larger values of $W$ here than those shown in Fig.~\ref{fig:LLL_map}, allowing the resonant oscillations to be more clearly separated from the nonresonant melting background.

For completeness, we also specify the spectral analysis used throughout this work. The spectra shown here, as well as those used to extract the dominant frequencies in the main text, are obtained using the Lomb--Scargle method rather than a direct discrete Fourier transform. For each time trace of $\langle S_0^z(t)\rangle$, the spectral power is evaluated on a dense frequency grid, and $\omega_{\mathrm{dom}}/2\pi$ is identified from its maximum. This provides a finer estimate of the dominant frequency than a standard discrete Fourier transform over a finite observation time.

\section{SIZE OF THE RESONANT SUBSPACE}
\label{app:subspace_size}

To further understand the remarkably long-lived coherent oscillations discussed in the main text, we examine the size of the resonant subspace associated with different initial states. For a given initial Fock state, we define $\mathcal{N}_{\mathrm{deg}}$ as the number of Fock configurations with the same diagonal energy as the initial state. Since the dynamical term in Hamiltonian~\eqref{eq:Hspin} consists of nearest-neighbor spin exchanges, all Fock configurations within a fixed total-magnetization sector can in principle be connected through sufficiently high-order exchange processes. We therefore take these energy-matched configurations to define the resonant subspace associated with the initial state, with dimension $\mathcal{N}_{\mathrm{deg}}$.

We compare two representative half-filled product states. The first is the single-domain-wall (DW) state $|\psi_0\rangle$ introduced in Eq.~\eqref{eq:initial_state} and studied throughout the main text. The second is a charge-density-wave (CDW) state with an alternating spin configuration,
\[
|\psi_{\mathrm{CDW}}\rangle
=
|\downarrow\uparrow\downarrow\uparrow\cdots\rangle .
\]

\begin{figure}[t]
    \centering
    \includegraphics[width=0.9\linewidth]{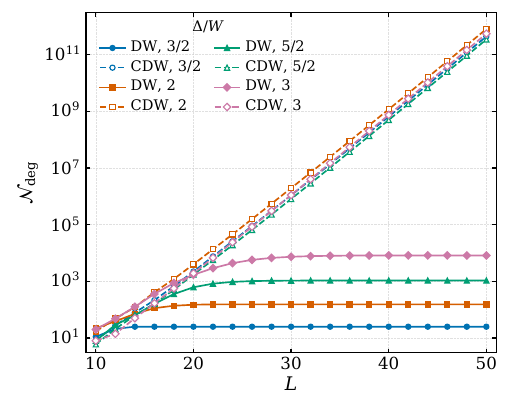}
    \caption{
    Size dependence of the resonant subspace.
    The dimension $\mathcal{N}_{\mathrm{deg}}$ of the resonant subspace is shown as a function of system size $L$ for resonance ratios $p/q=3/2$, $2$, $5/2$, and $3$.
    Solid lines correspond to the single-domain-wall (DW) initial state studied in the main text, while dashed lines correspond to the charge-density-wave (CDW) state.
    For the DW state, $\mathcal{N}_{\mathrm{deg}}$ saturates with increasing $L$, whereas for the CDW state it grows exponentially.
    }
    \label{fig:subspace_size}
\end{figure}

Figure~\ref{fig:subspace_size} shows a clear initial-state dependence of the resonant-subspace dimension. For the single-domain-wall state, $\mathcal{N}_{\mathrm{deg}}$ saturates with increasing $L$ for different resonance ratios considered, whereas for the CDW state it grows exponentially. This saturation for the DW state is crucial for the remarkably long-lived coherent oscillations discussed in the main text, since the resonant dynamics remain effectively confined to a finite-dimensional subspace as the system size increases. More importantly, the sharp contrast between the DW and CDW initial states demonstrates that the size of the resonant subspace depends crucially on the initial-state structure, rather than solely on the commensurate resonance condition $\Delta/W=p/q$. This strong initial-state dependence underlies the connection to the weak-fragmentation picture in Discussion~\cite{sala2020ergodicity,moudgalya2022quantum,wang2025exploring}.

\section{DETAILS OF THE EFFECTIVE HAMILTONIAN}
\label{app:effective_Hamiltonian}

\subsection{Derivation of the third-order effective Hamiltonian in operator form}

In the rotating frame, the spinless-fermion Hamiltonian is given by Eq.~\eqref{eq:Hrot} in the main text. According to the three possible values of $n_{i-1}-n_{i+2}$, we introduce the projection operators
\[
\begin{split}
    &P_i^{(0)}=(1-n_{i-1})(1-n_{i+2})+n_{i-1}n_{i+2},\\
    &P_i^{(+)}=n_{i-1}(1-n_{i+2}),\\
    &P_i^{(-)}=(1-n_{i-1})n_{i+2},
\end{split}
\]
which project onto the sectors with $n_{i-1}-n_{i+2}=0$, $+1$, and $-1$, respectively. The density-dependent phase factor in Eq.~\eqref{eq:Hrot} can then be decomposed as
\[
e^{iUt(n_{i-1}-n_{i+2})}
=
P_i^{(0)}
+e^{iUt}P_i^{(+)}
+e^{-iUt}P_i^{(-)}.
\]

At the third-order resonance $U/F=3$, Eq.~\eqref{eq:Hrot} therefore takes the explicit form
\begin{equation}
\begin{split}
    H^{\mathrm{rot}}(t)
    =
    \mathsf{t}\sum_i
    \Big(
    &e^{iFt}P_i^{(0)}c_{i+1}^{\dagger}c_i
    +e^{-i2Ft}P_i^{(+)}c_{i+1}^{\dagger}c_i\\
    &+e^{i4Ft}P_i^{(-)}c_{i+1}^{\dagger}c_i
    \Big)
    +\mathrm{h.c.}.
\end{split}
\label{eq:Hrot_explicit}
\end{equation}

Using the Fourier decomposition defined in Eq.~\eqref{eq:Fourier_component}, the six nonzero Fourier components are
\begin{equation}
\begin{split}
    H_1
    &=\mathsf{t}\sum_i P_i^{(0)}c_{i+1}^{\dagger}c_i
    \equiv \mathsf{t}\sum_i A_i,
    \qquad
    H_{-1}=\mathsf{t}\sum_i A_i^{\dagger},
    \\
    H_{-2}
    &=\mathsf{t}\sum_i P_i^{(+)}c_{i+1}^{\dagger}c_i
    \equiv \mathsf{t}\sum_i B_i,
    \qquad
    H_2=\mathsf{t}\sum_i B_i^{\dagger},
    \\
    H_4
    &=\mathsf{t}\sum_i P_i^{(-)}c_{i+1}^{\dagger}c_i
    \equiv \mathsf{t}\sum_i C_i,
    \qquad
    H_{-4}=\mathsf{t}\sum_i C_i^{\dagger}.
\end{split}
\label{eq:Fourier_components_explicit}
\end{equation}

Substituting these Fourier components into the van Vleck high-frequency expansion given in Eq.~\eqref{eq:HFE} of the main text, we obtain
\begin{equation}
\begin{split}
    &H_{\mathrm{eff}}^{(0)}=0,\\
    &H_{\mathrm{eff}}^{(1)}=\frac{1}{F}\Big(
    [H_1,H_{-1}]
    +\frac{1}{2}[H_2,H_{-2}]
    +\frac{1}{4}[H_4,H_{-4}]
    \Big),\\
    &H_{\mathrm{eff}}^{(2)}{=}\frac{1}{F^2}\Big(
    \frac{1}{2}[H_1,[H_1,H_{-2}]]
    {+}\frac{1}{8}[H_2,[H_2,H_{-4}]]
    \Big){+}\mathrm{h.c.}.
    \label{eq:HFE_explicit}
\end{split}
\end{equation}
The resulting effective Hamiltonian can then be obtained by evaluating these commutators using the canonical fermionic anticommutation relations
\begin{equation}
\begin{split}
    \{c_i^{\dagger},c_j\}&=\delta_{ij},
    \qquad
    \{c_i,c_j\}=0,
    \qquad
    \{c_i^{\dagger},c_j^{\dagger}\}=0,\\
    [n_i,c_j]&=-c_j\delta_{ij},
    \qquad
    [n_i,c_j^{\dagger}]=c_j^{\dagger}\delta_{ij}.
\end{split}
\end{equation}

The three commutators entering $H_{\mathrm{eff}}^{(1)}$ are
\[
\begin{split}
    [H_1,H_{-1}]
    =\mathsf{t}^2 &\sum_i \Big[
    P_i^{(0)}(n_{i+1}-n_i)\\
    &+(n_{i-1}-n_{i+4})
    \big(c_{i+1}^{\dagger}c_i
    c_{i+2}^{\dagger}c_{i+3}
    +\mathrm{h.c.}\big)
    \Big],
    \\
    [H_2,H_{-2}]
    =\mathsf{t}^2 &\sum_i \Big[
    -P_i^{(+)}(n_{i+1}-n_i)\\
    &+n_{i-1}(1-n_{i+4})
    \big(c_{i+1}^{\dagger}c_i
    c_{i+2}^{\dagger}c_{i+3}
    +\mathrm{h.c.}\big)
    \Big],
    \\
    [H_4,H_{-4}]
    =\mathsf{t}^2 &\sum_i \Big[
    P_i^{(-)}(n_{i+1}-n_i)\\
    &+(1-n_{i-1})n_{i+4}
    \big(c_{i+1}^{\dagger}c_i
    c_{i+2}^{\dagger}c_{i+3}
    +\mathrm{h.c.}\big)
    \Big].
\end{split}
\]

Combining these contributions gives
\begin{equation}
\begin{split}
    H_{\mathrm{eff}}^{(1)}
    =\frac{\mathsf{t}^2}{F}\sum_i
    &\left[
    P_i^{(0)}
    -\frac{1}{2}P_i^{(+)}
    +\frac{1}{4}P_i^{(-)}
    \right]
    (n_{i+1}-n_i)
    \\
    +\frac{\mathsf{t}^2}{F}\sum_i
    &\left[
    \frac{3}{2}n_{i-1}(1-n_{i+4})
    -\frac{3}{4}(1-n_{i-1})n_{i+4}
    \right]
    \\
    &\times
    \left(
    c_{i+1}^{\dagger}c_i
    c_{i+2}^{\dagger}c_{i+3}
    +\mathrm{h.c.}
    \right).
    \label{eq:Heff1_explicit}
\end{split}
\end{equation}

The first term in Eq.~\eqref{eq:Heff1_explicit} is a diagonal energy correction that depends only on the local occupations and therefore does not induce transitions between Fock configurations. The second term describes constrained second-order pair-hopping processes within the resonant subspace, which connects two different Fock states with the same diagonal energy.

For $H_{\mathrm{eff}}^{(2)}$, the nested commutators can be evaluated in the same manner. Collecting the resulting terms, we obtain nine distinct types of third-order dynamical processes,
\begin{equation}
\begin{split}
T_1&=\sum_i n_{i-1}(1-n_{i+1})(1-n_{i+2})(1-n_{i+4})
\,c_{i+3}^{\dagger}c_i,\\
T_2&=\sum_i n_{i-1}n_{i+1}n_{i+2}(1-n_{i+4})
\,c_{i+3}^{\dagger}c_i,\\
T_3&=\sum_i n_{i-1}(1-n_{i+3})(1-n_{i+5})
\,c_{i+1}^{\dagger}c_{i+4}^{\dagger}c_ic_{i+2},\\
T_4&=\sum_i (1-n_{i-1})n_{i+1}(1-n_{i+2})(1-n_{i+4})
\,c_{i+3}^{\dagger}c_i,\\
T_5&=\sum_i n_{i-1}(1-n_{i+1})(1-n_{i+5})
\,c_{i+2}^{\dagger}c_{i+4}^{\dagger}c_ic_{i+3},\\
T_6&=\sum_i n_{i-1}n_{i+3}(1-n_{i+5})
\,c_{i+1}^{\dagger}c_{i+4}^{\dagger}c_ic_{i+2},\\
T_7&=\sum_i n_{i-1}n_{i+1}(1-n_{i+2})n_{i+4}
\,c_{i+3}^{\dagger}c_i,\\
T_8&=\sum_i n_{i-1}n_{i+1}(1-n_{i+5})
\,c_{i+2}^{\dagger}c_{i+4}^{\dagger}c_ic_{i+3},\\
T_9&=\sum_i n_{i-1}(1-n_{i+6})
\,c_{i+1}^{\dagger}c_{i+3}^{\dagger}c_{i+5}^{\dagger}
c_ic_{i+2}c_{i+4}.
\end{split}
\label{eq:third_order_dynamical_explicit}
\end{equation}
Together with their Hermitian conjugates, these processes give
\begin{equation}
\begin{split}
H_{\mathrm{eff}}^{(2)}
=
\frac{\mathsf{t}^3}{F^2}
\Big(
&\frac{1}{2}T_1
+\frac{1}{2}T_2
+\frac{3}{2}T_3
-\frac{1}{4}T_4
+\frac{3}{4}T_5\\
&-\frac{3}{4}T_6
-\frac{1}{4}T_7
-\frac{3}{2}T_8
-\frac{9}{4}T_9
\Big)
+\mathrm{h.c.}.
\label{eq:Heff2_explicit}
\end{split}
\end{equation}

When acting on the single-domain-wall initial state~\eqref{eq:initial_state}, only the $T_1$ and $T_2$ process types act nontrivially. These are therefore the third-order processes directly responsible for the resonant dynamics discussed in the main text.

The effective Hamiltonian at the third-order resonance $U/F=3$, up to order $\mathsf{t}^3/F^2$, is thus
\begin{equation}
H_{\mathrm{eff}}
=
H_{\mathrm{eff}}^{(1)}
+
H_{\mathrm{eff}}^{(2)},
\label{eq:Heff_full_explicit}
\end{equation}
with $H_{\mathrm{eff}}^{(1)}$ and $H_{\mathrm{eff}}^{(2)}$ given by Eqs.~\eqref{eq:Heff1_explicit} and~\eqref{eq:Heff2_explicit}. The first-order-in-$1/F$ contribution contains both diagonal energy corrections and constrained second-order pair-hopping processes, while the second-order-in-$1/F$ contribution generates the third-order resonant processes. The effective-Hamiltonian dynamics shown in Fig.~\ref{fig:overlap_dynamics}(b) are obtained by quenching the same single-domain-wall initial state under Eq.~\eqref{eq:Heff_full_explicit}.

\subsection{Matrix-element representation of the effective Hamiltonian}

The operator form derived above explicitly identifies the microscopic processes generated by the high-frequency expansion. For higher orders, however, it can be more convenient to construct the effective Hamiltonian directly through matrix elements between resonant Fock states. This viewpoint is naturally provided by degenerate Schrieffer--Wolff perturbation theory~\cite{klein1974degenerate,bravyi2011schrieffer,bukov2016schrieffer,chen2025many}: virtual transitions through off-resonant Fock states generate effective couplings between states with the same diagonal energy. In the present problem, this construction is equivalent, order by order, to the van Vleck expansion used above.

To make this connection more explicit, we write the fermionic Hamiltonian as $H_{\mathrm{fermion}}=H_D+V$, where $H_D$ contains the diagonal Stark and interaction terms and $V$ is the nearest-neighbor hopping term. We label the diagonal-energy subspaces of $H_D$ by Greek indices $\alpha,\beta,\gamma$, with corresponding energies $E_\alpha,E_\beta,E_\gamma$, and denote an individual Fock state in subspace $\alpha$ by $|a,\alpha\rangle$, such that $H_D|a,\alpha\rangle=E_\alpha|a,\alpha\rangle$.

Consider two Fock states $|a,\alpha\rangle$ and $|b,\alpha\rangle$ belonging to the same resonant subspace. At second order, an effective coupling between them is generated by a two-step virtual process in which the system first hops to an intermediate state $|m,\beta\rangle$ with $\beta\neq\alpha$ and subsequently returns to the resonant subspace. The intermediate state is detuned from the resonant energy by $E_\beta-E_\alpha$, so standard degenerate perturbation theory weights this virtual path by the inverse energy mismatch $1/(E_\alpha-E_\beta)$. Summing over all possible intermediate states gives
\begin{equation}
\langle a,\alpha|
H_{\mathrm{eff}}^{(1)}
|b,\alpha\rangle
=
\sum_{\beta\neq\alpha}
\sum_{m\in\beta}
\frac{
\langle a,\alpha|V|m,\beta\rangle
\langle m,\beta|V|b,\alpha\rangle
}{
E_\alpha-E_\beta
},
\label{eq:Heff1_matrix_element}
\end{equation}
where the case $a=b$ gives diagonal energy corrections, while $a\neq b$ gives second-order off-diagonal couplings within the resonant subspace.

The connection with the van Vleck expansion follows directly from the Fourier decomposition of the rotating-frame Hamiltonian. A component $H_l$ collects single-hopping processes that change the diagonal energy by $lF$. For example, $H_1$ in Eq.~\eqref{eq:Fourier_components_explicit} moves a fermion one site to the right without changing the interaction energy and therefore changes the diagonal energy by $F$. Similarly, $H_{-2}$ describes a rightward hop accompanied by a decrease of the interaction energy by $U$, giving $F-U=-2F$ at $U/F=3$. Thus, the factors $1/(lF)$ appearing in the van Vleck expansion~\eqref{eq:HFE} are precisely the inverse energy mismatches associated with the virtual intermediate states in Eq.~\eqref{eq:Heff1_matrix_element}.

The same picture extends directly to third order. A three-step virtual process connecting two states in the resonant subspace passes through two intermediate states, $|m,\beta\rangle$ and $|n,\gamma\rangle$, and therefore carries two energy denominators. Summing over all such paths gives
\begin{equation}
\begin{split}
&\langle a,\alpha|
H_{\mathrm{eff}}^{(2)}
|b,\alpha\rangle\\
&=
\sum_{\substack{\beta,\gamma\neq\alpha\\
m\in\beta,\;n\in\gamma}}
\frac{
\langle a,\alpha|V|n,\gamma\rangle
\langle n,\gamma|V|m,\beta\rangle
\langle m,\beta|V|b,\alpha\rangle
}{
(E_\alpha-E_\gamma)(E_\alpha-E_\beta)
}.
\end{split}
\label{eq:Heff2_matrix_element}
\end{equation}
At the third-order resonance considered here, there is no zero-frequency hopping Fourier component, $H_0=0$, and hence no direct first-order hopping between Fock states within the same diagonal-energy subspace. As a result, the additional terms in the general third-order degenerate perturbation expansion that contain such in-subspace first-order matrix elements vanish, leaving the three-step virtual-path expression in Eq.~\eqref{eq:Heff2_matrix_element}.

Equations~\eqref{eq:Heff1_matrix_element} and~\eqref{eq:Heff2_matrix_element} reproduce the matrix elements of the operator expressions in Eqs.~\eqref{eq:Heff1_explicit} and~\eqref{eq:Heff2_explicit}. While the operator form makes the local structure of the resonant processes explicit, the matrix-element form provides a direct construction of the effective Hamiltonian within a chosen resonant subspace by summing the allowed virtual hopping paths and their corresponding energy denominators.

\subsection{Generalization to higher orders}

The path-sum construction introduced above extends straightforwardly to higher perturbative orders. As an example, consider the fourth-order resonance at $U/F=4$, corresponding to $\Delta/W=2$ in the original spin representation. In this case, the effective couplings arise in $H_{\mathrm{eff}}^{(3)}$ from sequences of four elementary hoppings. Although writing the complete fourth-order effective Hamiltonian in closed operator form requires evaluating considerably more complicated nested commutators, the corresponding matrix elements can be obtained directly from the virtual paths connecting resonant Fock states:
\begin{equation}
\begin{split}
&\langle a,\alpha|
H_{\mathrm{eff}}^{(3)}
|b,\alpha\rangle
\supset\\
&{\sum_{\substack{
\beta,\gamma,\theta\neq\alpha\\
m\in\beta\\
n\in\gamma\\
k\in\theta}}}
\frac{
\langle a{,}\alpha|V|k{,}\theta\rangle
\langle k{,}\theta|V|n{,}\gamma\rangle
\langle n{,}\gamma|V|m{,}\beta\rangle
\langle m{,}\beta|V|b{,}\alpha\rangle
}{
(E_\alpha-E_\theta)
(E_\alpha-E_\gamma)
(E_\alpha-E_\beta)
}.
\end{split}
\label{eq:Heff3_matrix_element}
\end{equation}
Each term represents a four-step virtual path connecting two Fock states in the same resonant subspace through three off-resonant intermediate states.

For the single-domain-wall initial state at $U/F=4$, the fourth-order effective dynamics connect the initial state to exactly three distinct resonant Fock configurations. The corresponding processes are
\begin{equation}
\begin{split}
T_a
&{=}
\sum_i
n_{i-1}
(1{-}n_{i+1})
(1{-}n_{i+2})
(1{-}n_{i+3})
(1{-}n_{i+5})
c_{i+4}^{\dagger}c_i{,}
\\
&T_a|1111100000\rangle
\propto
|1111000010\rangle,
\\[2mm]
T_b
&{=}
\sum_i
n_{i-1}
n_{i+1}
n_{i+2}
n_{i+3}
(1{-}n_{i+5})
c_{i+4}^{\dagger}c_i,
\\
&T_b|1111100000\rangle
\propto
|1011110000\rangle,
\\[2mm]
T_c
&{=}
\sum_i
n_{i-1}
(1{-}n_{i+4})
c_{i+2}^{\dagger}
c_{i+3}^{\dagger}
c_i c_{i+1},
\\
&T_c|1111100000\rangle
\propto
|1110011000\rangle.
\end{split}
\label{eq:fourth_order_dynamical}
\end{equation}

Although $T_a$, $T_b$, and $T_c$ are all fourth-order resonant processes, their effective couplings need not be identical because they involve different virtual paths and energy denominators. We denote the corresponding target Fock states by
\[
|\psi_\mu\rangle \propto T_\mu|\psi_0\rangle,
\qquad
\mu=a,b,c,
\]
and define the effective couplings as
\begin{equation}
g_\mu
=
\langle \psi_\mu|
H_{\mathrm{eff}}^{(3)}
|\psi_0\rangle .
\label{eq:fourth_order_couplings}
\end{equation}
Summing over the allowed four-step virtual paths in Eq.~\eqref{eq:Heff3_matrix_element}, we obtain
\begin{equation}
|g_a|=|g_b|,
\qquad
|g_c|=\frac{2}{5}|g_a|.
\label{eq:fourth_order_coupling_ratio}
\end{equation}
Thus, while all three channels share the fourth-order scaling $g_\mu\sim \mathsf{t}^4/F^3$, their relative strengths are determined by the detailed structure of the virtual paths.

To test this prediction dynamically, we perform exact-diagonalization time evolution for $L=20$ at $F=4$ and $U=16$, corresponding to the fourth-order resonance $U/F=4$. We monitor the projection probabilities
\begin{equation}
P_\mu(t)
=
|\langle\psi_\mu|\psi(t)\rangle|^2,
\qquad
\mu=0,a,b,c,
\label{eq:fourth_order_projection}
\end{equation}
where $|\psi_0\rangle$ is the initial single-domain-wall state and $|\psi_{a,b,c}\rangle$ are the three target Fock states defined above.

\begin{figure}[t]
    \centering
    \includegraphics[width=0.9\linewidth]{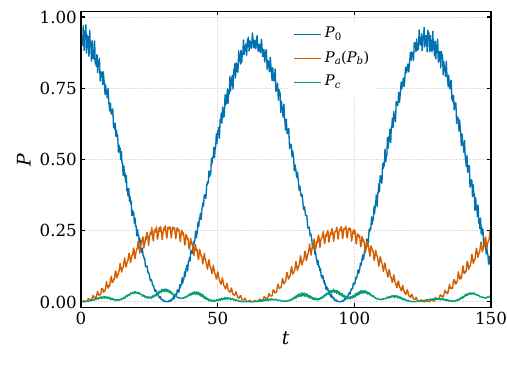}
    \caption{
    Projection dynamics at the fourth-order resonance. Numerical results are obtained by exact diagonalization of the full Hamiltonian for $L=20$, $F=4$, and $U=16$.
    $P_0$ denotes the projection probability onto the initial single-domain-wall state, while $P_a$, $P_b$, and $P_c$ denote those onto the target Fock states generated by $T_a$, $T_b$, and $T_c$, respectively.
    The curves $P_a$ and $P_b$ coincide, consistent with their equal effective couplings $|g_a|=|g_b|$, whereas $P_c$ remains substantially smaller, reflecting the weaker coupling $|g_c|=(2/5)|g_a|$.
    }
    \label{fig:fourth_order_projection}
\end{figure}

As shown in Fig.~\ref{fig:fourth_order_projection}, $P_a(t)$ and $P_b(t)$ coincide throughout the evolution, while $P_c(t)$ remains substantially smaller. This hierarchy is consistent with the coupling strengths in Eq.~\eqref{eq:fourth_order_coupling_ratio}: although all three target states are connected to the initial state by fourth-order resonant processes, the $T_c$ channel is significantly weaker.

This example demonstrates the utility of the matrix-element representation for higher-order resonances. The resonance condition identifies the resonant subspace, while the virtual-path summation determines the relative coupling strengths of individual resonant channels. At sufficiently high orders, identifying all allowed virtual paths can itself become combinatorially demanding; nevertheless, the path-based formulation remains systematic and often provides a more transparent route to the relevant effective couplings than deriving a complete closed operator form.

\bibliography{cite}

\end{document}